\documentclass[conference]{IEEEtran}
\IEEEoverridecommandlockouts
\usepackage{tikz}
\nonstopmode
\usepackage{cite}
\usepackage{amsmath,amssymb,amsfonts}
\usepackage[linesnumbered,lined,ruled,vlined]{algorithm2e}
\usepackage{graphicx}
\usepackage{textcomp}
\usepackage{xcolor}
\usepackage{hyperref}
\usepackage{cleveref}
\usepackage{commath}
\usepackage{optidef}
\usepackage[symbol]{footmisc}
\usepackage{lipsum}
\usepackage{enumitem}
\allowdisplaybreaks
\usepackage[left=1.62cm,right=1.62cm,top=1.7cm,bottom=2.67cm]{geometry}
\IEEEaftertitletext{\vspace{-0.5\baselineskip}}

\makeatletter
\newcommand{\nosemic}{\renewcommand{\@endalgocfline}{\relax}}
\newcommand{\dosemic}{\renewcommand{\@endalgocfline}{\algocf@endline}}

\let\oldnl\nl
\newcommand{\nonl}{\renewcommand{\nl}{\let\nl\oldnl}}
\makeatother

\def\BibTeX{{\rm B\kern-.05em{\sc i\kern-.025em b}\kern-.08em
    T\kern-.1667em\lower.7ex\hbox{E}\kern-.125emX}}

\begin{document}

\title{Deep Reinforcement Learning for Trajectory and Phase Shift Optimization of Aerial RIS in CoMP-NOMA Networks}
\author{
\IEEEauthorblockN{
    Muhammad Umer\IEEEauthorrefmark{1},
    Muhammad Ahmed Mohsin\IEEEauthorrefmark{1},
    Aamir Mahmood\IEEEauthorrefmark{2},
    Kapal Dev\IEEEauthorrefmark{3},
    Haejoon Jung\IEEEauthorrefmark{4},\\
    Mikael Gidlund\IEEEauthorrefmark{2},
    and Syed Ali Hassan\IEEEauthorrefmark{1}}
\IEEEauthorblockA{\IEEEauthorrefmark{1}School of Electrical Engineering and Computer Science (SEECS),\\National University of Sciences and Technology (NUST), Pakistan}
\IEEEauthorblockA{\IEEEauthorrefmark{2}Department of Computer and Electrical Engineering, Mid Sweden University, Sweden}
\IEEEauthorblockA{\IEEEauthorrefmark{3}Department of Computer Science, Munster Technological University, Ireland}
\IEEEauthorblockA{\IEEEauthorrefmark{4}Department of Electronics and Information Convergence Engineering, Kyung Hee University, Republic of Korea}
\IEEEauthorblockA{Email: \{mumer.bee20seecs, mmohsin.bee20seecs, ali.hassan\}@seecs.edu.pk,\\\{aamir.mahmood, mikael.gidlund\}@miun.se, kapal.dev@ieee.org, haejoonjung@khu.ac.kr
}
}

\maketitle

\begin{abstract}
    This paper explores the potential of aerial reconfigurable intelligent surfaces (ARIS) to enhance coordinated multi-point non-orthogonal multiple access (CoMP-NOMA) networks. We consider a system model where a UAV-mounted RIS assists in serving multiple users through NOMA while coordinating with multiple base stations. The optimization of UAV trajectory, RIS phase shifts, and NOMA power control constitutes a complex problem due to the hybrid nature of the parameters, involving both continuous and discrete values. To tackle this challenge, we propose a novel framework utilizing the multi-output proximal policy optimization (MO-PPO) algorithm. MO-PPO effectively handles the diverse nature of these optimization parameters, and through extensive simulations, we demonstrate its effectiveness in achieving near-optimal performance and adapting to dynamic environments. Our findings highlight the benefits of integrating ARIS in CoMP-NOMA networks for improved spectral efficiency and coverage in future wireless networks.

\end{abstract}


\section{Introduction}
Driven by the ever-growing demand for ubiquitous connectivity and high data rates, future wireless networks necessitate the exploration of novel solutions that surpass the limitations of traditional approaches. Unmanned aerial vehicles (UAVs), with their inherent mobility and flexible deployment as aerial base stations, have emerged as a promising technology to address these challenges. This enables them to provide wireless service in diverse scenarios, ranging from temporary hotspots during events and disaster-stricken areas with compromised infrastructure to remote locations with limited coverage~\cite{mohamed2020leveraging, mozaffari2019tutorial}. However, UAV-assisted networks face challenges such as limited energy capacity and constrained coverage area, hindering the full realization of their potential benefits.

To overcome these limitations and unlock the full potential of UAV-assisted networks, researchers are actively investigating the integration of enabling technologies such as reconfigurable intelligent surfaces (RIS)~\cite{gao2021aerial}. RIS are engineered surfaces composed of numerous passive reflecting elements that offer the ability to control electromagnetic wave propagation. By intelligently manipulating the phase shifts of the incident signals, RIS can enhance the desired signal strength, suppress interference, and extend coverage area. In the context of UAV-assisted networks, mounting RIS on UAVs creates aerial RIS (ARIS) networks, offering greater flexibility in optimizing the wireless environment through dynamic adaptation of RIS location and orientation~\cite{do2021aerial}. This dynamic adaptability enables ARIS to proactively respond to changing channel conditions and user distribution, fostering efficient and robust communication links.


Furthermore, non-orthogonal multiple access (NOMA) and coordinated multi-point (CoMP) transmission offer complementary benefits for improving spectral efficiency and user fairness in wireless networks. NOMA enables multiple users to share the same time-frequency resources, enhancing spectrum utilization and performance metrics such as outage probability and spectral efficiency~\cite{yue2018unified}. CoMP, on the other hand, enables cooperation between multiple base stations to jointly serve users, mitigating inter-cell interference and enhancing user experience. The integration of CoMP and NOMA (CoMP-NOMA) further amplifies these benefits by allowing multiple BSs to collaboratively serve NOMA users with coordinated power allocation and SIC decoding~\cite{ali2018downlink, elhattab2020comp}.

The convergence of RIS, CoMP, and NOMA within UAV-assisted networks holds immense potential for enhancing the performance and efficiency of future wireless communication systems. While recent research has demonstrated the benefits of combining these technologies with UAVs, existing works often assume static RIS deployments, limiting network adaptability~\cite{zhao2022ris, budhiraja2022energy}. Although some studies have investigated ARIS-assisted CoMP-NOMA networks and optimized UAV trajectory and RIS phase shifts for sum rate maximization~\cite{lv2023uav}, the employed optimization approaches, such as double-layer alternating optimization, may face scalability and convergence challenges in large-scale networks.


To address the challenges of optimizing ARIS-assisted CoMP-NOMA networks, we propose a deep reinforcement learning (DRL) approach based on the multi-output proximal policy optimization (MO-PPO) algorithm to effectively address the hybrid continuous-discrete action space inherent in these networks. Our framework jointly optimizes UAV trajectory, RIS phase shifts, and NOMA power control to maximize network sum rate while adhering to user quality of service (QoS) constraints. Through extensive simulations, we evaluate the efficacy of our proposed approach, assess the convergence of MO-PPO, and highlight the benefits of CoMP-NOMA and RIS in UAV-assisted networks.


\section[System Model and Problem Formulation]{System Model \& Problem Formulation}

\begin{figure}
    \centering
    \includegraphics[width=0.84\columnwidth]{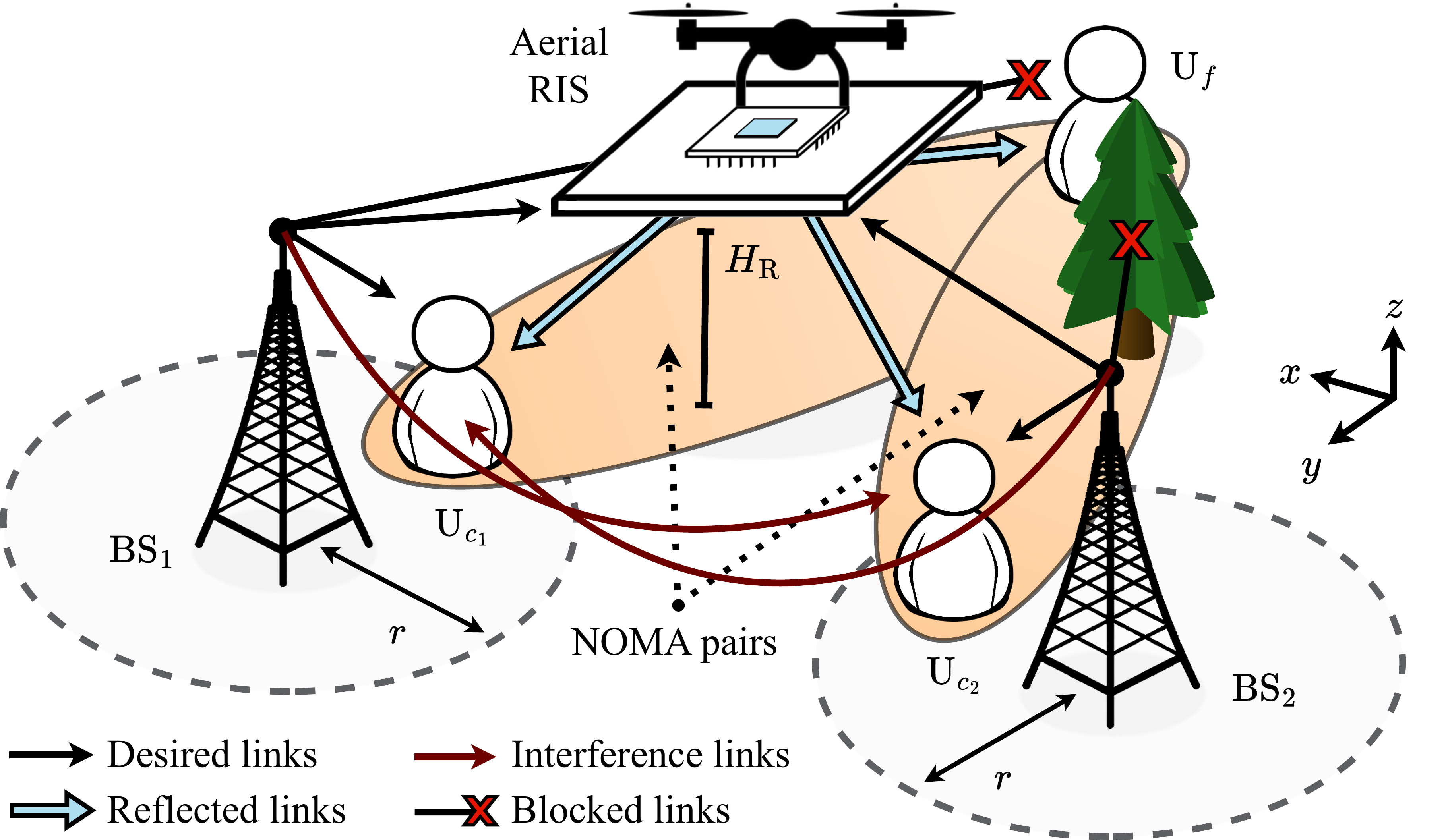}
    \caption{Aerial RIS-assisted coordinated NOMA cluster.}
    \label{fig:model}
    \vspace*{-1em}
\end{figure}

\subsection{System Description}
We consider a downlink transmission scenario in a multi-cell CoMP-NOMA network assisted by a UAV-mounted RIS, as illustrated in Fig~\ref{fig:model}. The network consists of $I$ cells, each modeled as a circular disk of radius $r$ with a single-antenna BS at its center, denoted as BS$_i$, where $i \in \mathcal{I} \triangleq \{1, 2, \ldots, I\}$. Each BS$_i$ invokes two-user downlink NOMA to serve its respective cell-center and edge user, each also equipped with a single-antenna. The cell-center users are defined as users that lie within the disk of their associated cell and are denoted as U$_{c_i}$, $\forall i$ and $c_i \in \mathcal{C}^i \triangleq \{1, 2, \ldots, C_i\}$, where $C_i$ is the number of cell-center users in cell $i$. Conversely, the edge users are defined as users that do not lie within any cell and are denoted as U$_{f}$, $\forall f$ and $f \in \mathcal{F} \triangleq \{1, 2, \ldots, F\}$, where $F$ is the number of edge users in the network. Furthermore, let $\mathcal{U} \triangleq \bigcup_{i \in \mathcal{I}} \mathcal{C}^i \cup \mathcal{F}$ be the set of all the users in the network. Without loss of generality and for ease of exposition, we assume $I = 2$, and $C_i = F = 1$, $\forall i$.

For coordinated operation, the BSs are assumed to be interconnected via a high-speed backhaul network to a central processing unit (CPU). Moreover, to improve the signal quality for edge users, an ARIS, denoted as R, is deployed at a fixed altitude $H_{\text{R}}$ over area $A$ to create reflection links between the BSs and the users, and is equipped with $K$ passive elements. For tractability, we discretize the entire system operation into time slots of equal length $\tau$, where each time slot is indexed by $t \in \mathcal{T} \triangleq \{1, 2, \ldots, T\}$, such that $T$ is the total flight time of the UAV. Furthermore, we assume the presence of $O$ obstacles in the network, denoted as $\mathcal{O} \triangleq \{1, 2, \ldots, O\}$, where each obstacle O$_o$, $o \in \mathcal{O}$ has its own \textit{forbidden zone} represented as a circular disk of radius $d_{\text{min}}$, centered at the obstacle's location, where the UAV is not allowed to fly due to safety and regulatory constraints.

Before proceeding with the channel and signal model, we define the positions of the various entities in the network. Specifically, $\forall i \in \mathcal{I}$, $u \in \mathcal{U}$, and $o \in \mathcal{O}$, the positions of BS$_i$, U$_u$, and O$_o$ are represented by $\mathbf{p}_{i}= (x_i, y_i, H_{\text{B}})$, $\mathbf{p}_{u}= (x_u, y_u, 0)$, and $\mathbf{p}_{o}= (x_o, y_o, H_{\text{O}})$, respectively, where $H_{\text{B}}$ and $H_{\text{O}}$ are the heights of the BSs and obstacles, respectively. Moreover, the position of R at time slot $t$ is denoted as $\mathbf{p}_{\text{R}}[t] = (x_{\text{R}}[t], y_{\text{R}}[t], H_{\text{R}})$. In this paper, we assume that the users are stationary, and the UAV is capable of adjusting its horizontal position in the $xy$-plane, while maintaining a fixed altitude.

\subsection{Channel Model \& RIS Configuration}
Our analysis considers both large-scale path loss and small-scale fading effects on signal propagation. Similar to~\cite{zhao2022ris}, we assume a rich scattering environment, leading to the modeling of direct links between BS$_i$ and U$_u$ as Rayleigh fading channels, denoted as $h_{i,u}$. Mathematically, the channel $h_{i,u}$ at time slot $t$ is given by
\begin{equation}
    h_{i, u}[t] = \sqrt{\frac{\rho_o}{PL(d_{i, u})}} v_{i, u}[t],
\end{equation}
where $\rho_o$ is the reference path loss at $1$ m, $PL(d_{i, u}) = (d_{i, u})^{\alpha_{i, u}}$ is the large-scale path loss, such that $\alpha_{i, u}$ is the path loss exponent, $d_{i, u}= \norm{\mathbf{p}_i - \mathbf{p}_u}$ is the distance between BS$_i$ and U$_u$ and $\norm{\cdot}$ denotes the Euclidean norm. Moreover, $v_{i, u}[t] \in \mathbb{C}^{1\times 1}$ is the small-scale Rayleigh fading coefficient with zero mean and unit variance, and is assumed to be independent and identically distributed (i.i.d) across different time slots and users. In this work, as a special case, we assume that the direct link between BS$_i$ and U$_f$ is blocked due to the presence of obstacles, thus $h_{i, f}[t] = 0$, $\forall i, f$.

Contrary to the direct links, the reflection links between BS$_i$ and R are modeled as Rician fading channels, denoted as $\mathbf{h}_{i, \text{R}}[t]$, due to the presence of a dominant line-of-sight (LoS) component. At time slot $t$, the channel $\mathbf{h}_{i, \text{R}}[t]$ is given by
\begin{multline}
    \mathbf{h}_{i, \text{R}}[t] = \\
    \sqrt{\frac{\rho_o}{PL(d_{i, \text{R}}[t])}} \left(\sqrt{\frac{\kappa}{1 + \kappa}} \mathbf{g}^{\text{LoS}}_{i, \text{R}}[t] + \sqrt{\frac{1}{1 + \kappa}} \mathbf{g}^{\text{NLoS}}_{i, \text{R}}[t]\right),
\end{multline}
where $\kappa$ is the Rician factor, and $d_{i, \text{R}}[t] = \norm{\mathbf{p}_i - \mathbf{p}_{\text{R}}[t]}$ is the distance between BS$_i$ and R. Moreover, the deterministic LoS represented, i.e., $\mathbf{g}_{i, \text{R}}^{\text{LoS}}[t] \in \mathbb{C}^{K \times 1}$, is given by
\begin{equation*}
    \mathbf{g}^{\text{LoS}}_{i, \text{R}} = \left[1, \ldots, e^{j(k-1)\pi\sin(\omega_i)}, \ldots, e^{j(K-1)\pi\sin(\omega_i)}\right]^T,
\end{equation*}
where $k \in \mathcal{K} \triangleq \{1, 2, \ldots, K\}$ indexes elements of R and $\omega_i$ is the angle of arrival (AoA) whereas $\mathbf{g}^{\text{NLoS}}_{i, \text{R}_i} \in \mathbb{C}^{K\times 1}$ is the NLoS component following Rayleigh fading as previously described. Similarly, the channel between R and U$_u$, denoted as $h_{\text{R}, u}$, can also be modeled as a Rician fading channel.

For the ARIS configuration, we assume that the phase shift of the $k$-th element can be set independently of other elements and that both the UAV trajectory and the phase response are controlled by the CPU. Furthermore, the phase shift (PS) matrix at time slot $t$ is expressed as
\begin{equation}
    \mathbf{\Theta}[t] = \text{diag}\left(a_1 e^{j\theta_1[t]}, a_2 e^{j\theta_2[t]}, \ldots, a_k e^{j\theta_K[t]}\right),
\end{equation}
where $a_k \in (0, 1]$ is the amplitude coefficient and $\theta_k[t] \in [-\pi, \pi)$ is the phase shift of the $k$-th element. For simplicity, we assume an ideal RIS with perfect phase control and unit amplitude reflection coefficients for all elements, i.e., $a_k = 1, \forall k$. Additionally, perfect channel state information (CSI) is assumed to be available at the CPU.

\subsection{Signal Model}
In accordance with the NOMA principle, each BS$_i$ serves two users, U$_{c_i}$ and U$_f$, simultaneously, by superimposing their signals. The transmitted signal from BS$_i$ at time slot $t$ is given by $x_i[t] = \sqrt{(1 - \lambda_i)P_i}x_{i, c_i}[t] + \sqrt{\lambda_i P_i}x_{i, f}[t]$, where $x_{i, c_i}[t]$ and $x_{i, f}[t]$ represent the respective signals for U$_{c_i}$ and U$_f$, $P_i$ is the transmit power of BS$_i$ and $\lambda_i$ is the power allocation factor assigned to U$_f$. To ensure successful decoding at the U$_{c_i}$, we constrain $\lambda_i \in (0.5, 1)$, as deduced in~\cite{elhattab2022ris, 9682500}.

The signal received by U$_{f}$ can be expressed as
\begin{equation}
    y_{f}[t] = H_{i, f}[t] x_i[t] + H_{i', f}[t] x_{i'}[t] + n_o[t]
\end{equation}
where $i' \in \mathcal{I} \setminus \{i\}$, $n_o[t] \sim \mathcal{CN}(0, \sigma^2)$ is the additive white Gaussian noise (AWGN), and $H_{i, f}[t] = \mathbf{h}^T_{\text{R}, f}[t] \mathbf{\Theta}[t] \mathbf{h}_{i, \text{R}}[t]$ represents the effective channels between BS$_i$ and U$_f$ through R, respectively. To minimize synchronization overhead, we employ non-coherent JT-CoMP, thus, the signal-to-interference-plus-noise ratio (SINR) is given by
\begin{equation}
    \gamma_f[t] = \frac{\lambda_i \abs{H_{i, f}[t]}^2 + \lambda_{i'} \abs{H_{i', f}[t]}^2}{(1 - \lambda_i) \abs{H_{i, f}[t]}^2 + (1 - \lambda_{i'}) \abs{H_{i', f}[t]}^2 + \frac{1}{\rho}},
\end{equation}
where $\rho = P_t/\sigma^2$ is the transmit SNR and $P_t = P_i$, $\forall i$ is the transmit power of each BS.

On the other hand, the signal received by U$_{c_i}$ can be expressed as
\begin{equation}
    y_{c_i}[t] = H_{i, c_i}[t] x_i[t] + h_{i', c_i}[t] x_{i'}[t] + n_o[t],
\end{equation}
where $H_{i, c_i}[t] = h_{i, c_i}[t] + \mathbf{h}^T_{\text{R}, c_i}[t] \mathbf{\Theta}[t] \mathbf{h}_{i, \text{R}}[t]$ represents the effective channels between BS$_i$ and U$_{c_i}$ through R, respectively. Also, the term $h_{i', c_i}[t] x_{i'}[t]$ represents the ICI caused by the transmission of BS$_{i'}$ at U$_{c_i}$. Based on the SIC principle, U$_{c_i}$ first decodes $x_{i, f}[t]$ and then cancels it from $y_{c_i}[t]$ to decode $x_{i, c_i}[t]$. The SINR at U$_{c_i}$ for decoding $x_{i, f}[t]$ is given by
\begin{equation}
    \gamma_{c_i \rightarrow f}[t] = \frac{\lambda_i \abs{H_{i, c_i}[t]}^2}{(1 - \lambda_i) \abs{H_{i, c_i}[t]}^2 + \abs{h_{i', c_i}[t]}^2 + \frac{1}{\rho}},
\end{equation}
whereas the SINR at U$_{c_i}$ for decoding $x_{i, c_i}[t]$ is
\begin{equation}
    \gamma_{c_i}[t] = \frac{(1 - \lambda_i) \abs{H_{i, c_i}[t]}^2}{\abs{h_{i', c_i}[t]}^2 + \frac{1}{\rho}}.
\end{equation}
Finally, the achievable sum rate of the network at time slot $t$ can be expressed as
\begin{equation}
    R_{\text{sum}}[t] = \sum_{i \in \mathcal{I}} R_{c_i}[t] + \sum_{f \in \mathcal{F}} R_f[t].
\end{equation}
where $R_{c_i}[t]=\log_2(1 + \gamma_{c_i}[t])$ and $R_f[t]=\log_2(1 + \gamma_f[t])$ are the achievable rates of U$_{c_i}$ and U$_f$, respectively.

\subsection{Problem Formulation}
The primary objective of this work is to maximize the cumulative sum rate achieved over a period of $T$ time slots. To accomplish this, we jointly optimize three key control variables: the UAV trajectory denoted as $\mathbf{P} \triangleq \{\mathbf{p}_{\text{R}}[t], \forall t\}$, the RIS phase shifts represented by $\mathbf{\Theta} \triangleq \{\mathbf{\Theta}[t], \forall t\}$, and the power allocation factors denoted as $\mathbf{\Lambda} \triangleq \{\lambda_i, \forall i\}$. Mathematically, the optimization problem is formulated as
\begin{maxi!}|s|
{\mathbf{P}, \mathbf{\Theta}, \mathbf{\Lambda}}{\sum_{t \in \mathcal{T}} R_{\text{sum}}[t]}{\label{eq:problem}}{}
\addConstraint{x_{\text{R}}[t], y_{\text{R}}[t] \in A,~\forall t \in \mathcal{T} \label{eq:loc_csrt}}
\addConstraint{\norm{\mathbf{p}_{\text{R}}[t] - \mathbf{p}_{o}} \geq d_{\text{min}},~\forall o \in \mathcal{O},~t \in \mathcal{T} \label{eq:obs_csrt}}
\addConstraint{\theta_k[t] \in [-\pi, \pi),~\forall k \in \mathcal{K},~t \in \mathcal{T} \label{eq:phase_csrt}}
\addConstraint{R_{c_i}[t] \geq R_{c_i}^{\text{min}},~\forall i \in \mathcal{I},~t \in \mathcal{T} \label{eq:ratec_csrt}}
\addConstraint{R_{f}[t] \geq R_{f}^{\text{min}},~\forall f \in \mathcal{F},~t \in \mathcal{T} \label{eq:ratef_csrt}}
\addConstraint{\lambda_i \in (0.5, 1),~\forall i \in \mathcal{I} \label{eq:power_csrt}},
\end{maxi!}
where constraint~\eqref{eq:loc_csrt} restricts the UAV trajectory to lie within $A$, and constraint~\eqref{eq:obs_csrt} enforces a minimum safety distance between the UAV and any obstacles present, thus guaranteeing the UAV's safety. Constraint~\eqref{eq:phase_csrt} limits the phase shifts applied by the RIS elements. To meet the quality of service (QoS) requirements, constraints~\eqref{eq:ratec_csrt} and~\eqref{eq:ratef_csrt} impose minimum rate thresholds, denoted by $\mathcal{R}_{c_i}^{\text{min}}$ and $\mathcal{R}_{f}^{\text{min}}$, for U$_{c_i}$ and U$_f$, respectively. Lastly, constraint~\eqref{eq:power_csrt} defines the permissible range for power allocation factors, ensuring successful SIC. The optimization problem in~\eqref{eq:problem} is non-convex due to the coupled variables $\{\mathbf{P}, \mathbf{\Theta}, \mathbf{\Lambda}\}$. To address this, we propose a DRL-based solution in the next section.

\section[DRL-based Proposed Solution]{Deep Reinforcement Learning-based \\Proposed Solution}

\subsection{MDP Formulation}
To enable the applicability of DRL, we first recast the problem in~\eqref{eq:problem} as a single-agent Markov Decision Process (MDP) operating in discrete time steps. The MDP is represented by the tuple $\langle \mathcal{S}, \mathcal{A}, \mathcal{P}, \mathcal{R}, \gamma \rangle$, where $\mathcal{S}$ denotes the set of possible environment states, $\mathcal{A}$ represents the action space, $\mathcal{P}$ defines the state transition probabilities, $\mathcal{R}$ is the reward function guiding the agent's learning, and $\gamma$ is the discount factor that determines the importance of future rewards. At each time slot $t$, the agent observes the current state $s_t$, selects an action $a_t$ based on its policy, transitions to a new state $s_{t+1}$, and receives a reward $\mathcal{R}(s_t, a_t)$. We define $\mathcal{S}$, $\mathcal{A}$, and $\mathcal{R}$ as follows.
\begin{enumerate}[wide=\parindent]
    \item \textit{State Space $\mathcal{S}$:} The environment state at time slot $t$ consists of the UAV's current position $\mathbf{p}_{\text{R}}[t]$, its distances to the center of each obstacle $\mathbf{d}_{\text{R}}[t] = \{\norm{\mathbf{p}_{\text{R}}[t] - \mathbf{p}_{o}}, \forall o \in \mathcal{O}\}$, the power allocation factors $\mathbf{\Lambda}$, and the achievable rates $\mathbf{R}[t] = \{R_{c_i}[t], R_f[t], \forall i, f\}$. Formally, the state space can be expressed as
          \begin{equation}
              s_t = \{\mathbf{p}_{\text{R}}[t], \mathbf{d}_{\text{R}}[t], \mathbf{\Lambda}, \mathbf{R}[t]\} \in \mathbb{R}^{\text{dim}_\mathcal{S}}.
          \end{equation}
          where $\text{dim}_\mathcal{S}=2 + O + I + \sum_{i \in \mathcal{I}} C_i + F$ is the dimension of the state space.

    \item \textit{Action Space $\mathcal{A}$:} The actions available to the agent consist of controlling the movement of the UAV in the $xy$-plane, adjusting the phase shifts of the RIS elements, and managing the power allocation factors. Specifically, the action space at time slot $t$ contains the manuevering actions $a_{\text{R}}[t] \in \{(-1, 0), (1, 0), (0, -1), (0, 1), (0, 0)\}$, representing left, right, down, up, and hover, respectively, the phase shifts $a_{\phi}[t] = \{\phi_k[t], \forall k\}$, and the power allocation factors $a_{\lambda} = \{\lambda_i, \forall i\}$. Thus, the action space can be represented as
          \begin{equation}
              a_t = \{a_{\text{R}}[t], a_{\phi}[t], a_{\lambda}\} \in \mathbb{R}^{\text{dim}_\mathcal{A}}.
          \end{equation}
          where $\text{dim}_\mathcal{A}=2 + K + I$ is the dimension of the action space.

    \item \textit{Reward Function $\mathcal{R}$:} The reward function plays a crucial role in shaping the learning behavior of the RL agent. Our design prioritizes maximizing the sum rate while simultaneously ensuring the UAV's safety and adherence to QoS requirements. The reward function is defined as
          \begin{equation}
              \label{eq:reward}
              \mathcal{R}(s_t, a_t) = R_{\text{sum}}[t] \left(1 - \frac{\sum_{u \in \mathcal{U}} \zeta_u[t]}{|\mathcal{U}|}\right) - \xi_{\text{R}}[t] K_{\text{viol}},
          \end{equation}
          where $K_{\text{viol}}$ is the penalty factor for UAV's safety constraint violation, and $\zeta_u[t]=\mathbb{I}\{R_u[t] \leq R_u^{\text{min}}\}$ is the indicator function for the QoS constraints, i.e., $\zeta_u[t]=1$ if QoS constraints are violated, and $0$ otherwise. Similarly, $\xi_{\text{R}}[t]=\mathbb{I}\{x_{\text{R}}[t], y_{\text{R}}[t] \notin A \land \norm{\mathbf{p}_{\text{R}}[t] - \mathbf{p}_{o}} < d_{\text{min}}, \forall o \in \mathcal{O}\}$ is the indicator function for UAV's safety constraints.
\end{enumerate}

\subsection{MO-PPO Algorithm}
The considered action space is a hybrid continuous-discrete space, which poses a challenge for traditional RL algorithms. While discretizing the continuous actions is a possible solution, it can lead to a large action space, significantly increasing computational complexity and potentially hindering performance. To address this challenge, we propose employing a multi-output Proximal Policy Optimization (MO-PPO) algorithm. MO-PPO extends the standard PPO~\cite{truly_ppo} framework by employing two parallel actor networks, each responsible for generating the discrete action $a_{\text{R}}$ and the continuous actions $a_{\phi}$ and $a_{\lambda}$, respectively. The actor networks share the first few layers, allowing for the extraction of common features and encoding state information. Furthermore, a single critic network is employed to estimate the value function $V(s_t)$, which is used to compute a variance-reduced advantage function estimate $\hat{A}_t$ for policy optimization. Following the implementation details used in~\cite{mnih2016asynchronous}, the policy is executed for $\hat{T}$ time steps, and $\hat{A}_t$ is computed as
\begin{equation}
    \label{eq:adv}
    \hat{A}_t = \sum_{k=0}^{\hat{T}-1} \gamma^k r_{t+k} + \gamma^{\hat{T}} V(s_{t+\hat{T}}) - V(s_t),
\end{equation}
where $\hat{T}$ is much smaller than the length of the episode $T$.

The discrete actor network outputs $|a_{\text{R}}|$ logits to generate the stochastic policy $\pi_{\theta_d}(a_t|s_t)$ for the discrete actions. These logits are passed through a $\mathrm{softmax}$ function to obtain a probability distribution over the discrete actions. Conversely, the continuous actor network generates $a_{\phi}$ and $a_{\lambda}$ by sampling from Gaussian distributions parameterized by the network's output means and standard deviations, as dictated by the stochastic policy $\pi_{\theta_c}(a_t|s_t)$. Both $\pi_{\theta_d}(a_t|s_t)$ and $\pi_{\theta_c}(a_t|s_t)$ are independently optimized using their respective clipped surrogate objective functions. The clipped surrogate objective function for the discrete actions is given by
\begin{equation}
    \label{eq:clip}
    L_d^{\text{CLIP}}(\theta_d) = \hat{\mathbb{E}}_t \left[\min(r_t^d(\theta_d) \hat{A}_t, \Im(r_t^d,\theta_d, \epsilon) \hat{A}_t\right],
\end{equation}
where $\Im(r_t^d,\theta_d, \epsilon) = \text{clip}(r_t^d(\theta_d), 1 - \epsilon, 1 + \epsilon)$, $r_t^d(\theta_d) = \pi_{\theta_d}(a_t|s_t)/\pi_{\theta_d}^{\text{old}}(a_t|s_t)$ is the importance sampling ratio, and $\epsilon$ is the clipping parameter. The objective function for the continuous actions can be expressed similarly but is omitted for brevity.

{\setlength{\algomargin}{1.25em}
    \begin{algorithm}[h!]
        \DontPrintSemicolon
        \caption{MO-PPO Algorithm}\label{alg:moppo}
        Initialize the policy parameters $\theta_d$ and $\theta_c$\;
        \For{episode $= 1, 2, \ldots, N$}{
        Receive initial state $s_0$\;
        \For{time step $t = 0, 1, \ldots, T$}{
            Generate discrete action $a_{\text{R}}$ using $\pi_{\theta_d}(a_t|s_t)$\;
            Generate continuous actions $a_{\phi}$ and $a_{\lambda}$\ using $\pi_{\theta_c}(a_t|s_t)$\;
            Execute actions $a_t = \{a_{\text{R}}, a_{\phi}, a_{\lambda}\}$\;
            \If{UAV violates~\eqref{eq:loc_csrt} or~\eqref{eq:obs_csrt}}{
                Set $\xi_{\text{R}}[t] = 1$, cancel the UAV's movement, and update the state $s_{t+1}$\;
            }
            Observe reward $\mathcal{R}$ as~\eqref{eq:reward} and next state $s_{t+1}$\;
            Collect a set of partial trajectories $\mathcal{D}$ with $\hat{T}$ transitions\;
            Compute the variance-reduced advantage estimate $\hat{A}_t$ as~\eqref{eq:adv}\;
        }
        \For{epoch $= 1, 2, \ldots, E$}{
        Sample a mini-batch of transitions $B$ from $\mathcal{D}$\;
        Compute the clipped surrogate objectives $L_d^{\text{CLIP}}(\theta_d)$ and $L_c^{\text{CLIP}}(\theta_c)$ as~\eqref{eq:clip}\;
        Optimize overall objective and update the policy parameters $\theta_d$ and $\theta_c$ using Adam~\;
        }
        Synchronize the sampling policies as\;
        \centerline{$\theta_d^{\text{old}} \leftarrow \theta_d$ and $\theta_c^{\text{old}} \leftarrow \theta_c$}
        Clear the collected trajectories $\mathcal{D}$\;
        }
    \end{algorithm}
}

Although both policies collaborate within the environment, their optimization objectives remain decoupled. This means $\pi_{\theta_d}(a_t|s_t)$ and $\pi_{\theta_c}(a_t|s_t)$ are treated as independent distributions during policy optimization, rather than a joint distribution encompassing both action spaces. The MO-PPO algorithm is summarized in Algorithm~\ref{alg:moppo}.

\subsection{Complexity and Convergence Analysis}
The complexity of DRL algorithms is commonly measured by the number of multiplications per iteration, which is a function of the number of parameters in the policy and value networks. For MO-PPO, the overall complexity can be expressed as $\mathcal{O}[\sum_{q=1}^{Q_s} n_q \cdot n_{q-1} + \sum_{q=1}^{Q_d} n_q \cdot n_{q-1} + \sum_{q=1}^{Q_c} n_q \cdot n_{q-1}]$, where $Q_s$, $Q_d$, and $Q_c$ are the number of layers in the shared, discrete, and continuous actor networks, respectively, and $n_q$ and $n_{q-1}$ are the number of neurons in the $q$-th and $(q-1)$-th layers, respectively. Assuming the same number of neurons in each hidden layer ($n_q = n_{q-1} = n$, $\forall q$) and an output layer size equal to the action space dimension, the overall complexity of MO-PPO simplifies to $\mathcal{O}[n^2(Q_s + Q_d + Q_c)]$.

Mathematically analyzing MO-PPO convergence is challenging due to the inherent complexities of neural networks and their dependence on hyperparameters. However, we can empirically verify convergence by monitoring the agent's performance over multiple episodes and ensuring that the reward function converges to a stable value.  Moreover, convergence can be accelerated by tuning the learning rate, clipping parameter, and penalty factor for constraint violation.

\section{Numerical Results}
\begin{table}[t!]
    \centering
    \caption{Simulation Parameters}
    \label{tab:params}
    \resizebox{0.99\columnwidth}{!}{%
        \begin{tabular}{|c|c|c|c|}
            \hline
            \textbf{Parameter}                      & \textbf{Value} & \textbf{Parameter}            & \textbf{Value}   \\
            \hline
            \hline
            Reference path loss $\rho_o$            & $-30$ dBm      & Rician factor $\kappa$        & $3$ dB           \\
            Target data rate $R_{f}^{\text{min}}$   & $0.2$ bps/Hz   & Learning rate                 & $2.75\text{e}-4$ \\
            Target data rate $R_{c_i}^{\text{min}}$ & $0.5$ bps/Hz   & Clipping parameter $\epsilon$ & $0.1$            \\
            Penalty constant $K_{\text{viol}}$      & $7$            & Discount factor $\gamma$      & $0.98$           \\
            Minimum distance $d_{\text{min}}$       & $10$ m         & Number of episodes $N$        & $750$            \\
            Time slots per episode $T$              & $250$          & Number of epochs $E$          & $20$             \\
            Number of neurons                       & $64$           & Batch size $B$                & $128$            \\
            \hline
        \end{tabular}%
    }
\end{table}

\subsection{Simulation Setup}
To evaluate the proposed MO-PPO algorithm, we construct a simulated urban environment spanning an area of $150 \times 150$ m$^2$ with $I = 2$ BSs, $U = 3$ users, and $O = 2$ obstacles. The initial position of the UAV is set to $(0, 35, 50)$ m, while BS$_1$ and BS$_2$ are located at $(-35, -35, 25)$ m and $(35, 35, 25)$ m, respectively. All remaining entities are randomly placed within the environment. Both BSs transmit at an identical power level, i.e., $P_1 = P_2 = P_t$. The network operates at a carrier frequency of $f_c = 2.4$ GHz, utilizing a bandwidth of $BW = 10$ MHz, and the noise power is set to $\sigma^2 = -174 + 10\log_{10}(BW)$ dBm. We model signal propagation using path loss exponents of $\alpha_{i, u} = 3$, $\alpha_{i, \text{R}} = \alpha_{\text{R}, u} = 2.2$, and $\alpha_{i', u} = 3.5$, for direct, reflection, and interference links, respectively. Table~\ref{tab:params} summarizes the remaining simulation parameters.

\subsection{Results}

\begin{figure}[t]
    \centerline{
        \includegraphics[width=0.75\columnwidth]{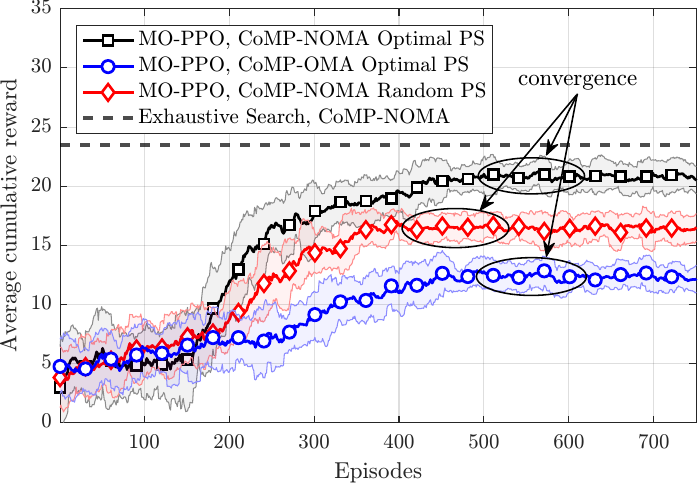}}
    \caption{Average cumulative reward vs. number of training episodes with $P_t = 20$ dBm and $K = 120$ elements.}
    \label{fig:reward}
\end{figure}

Fig~\ref{fig:reward} shows the average cumulative reward achieved by the MO-PPO algorithm under different network configurations. The algorithm consistently converges to a stable reward value after approximately 500 episodes, indicating the successful learning of an effective policy. Notably, MO-PPO with random phase shifts (PS) exhibits faster convergence than its counterpart with optimal PS. This can be attributed to the increased complexity associated with optimizing PS, leading to slower convergence. The CoMP-NOMA configuration also yields a higher average cumulative reward compared to the CoMP-OMA configuration, highlighting the benefits of NOMA in enhancing network performance. A comparison with the optimal solution obtained through exhaustive search reveals that the proposed MO-PPO algorithm achieves near-optimal performance, demonstrating its effectiveness.

\begin{figure}[t]
    \centerline{
        \includegraphics[width=0.75\columnwidth]{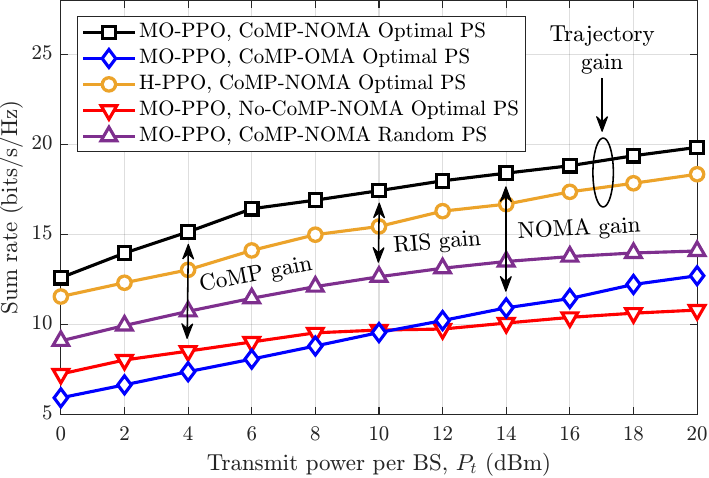}}
    \caption{Sum rate vs. transmit power for different algorithms and configurations with $K = 120$ elements.}
    \label{fig:rate_pt}
\end{figure}

Next, we investigate the sum rate of the network as a function of transmit power ($P_t$) as shown in Fig.~\ref{fig:rate_pt}. As expected, the sum rate increases with transmit power, highlighting the importance of power control in network optimization. The results showcase the advantages of integrating CoMP, RIS, and NOMA for enhanced spectral efficiency. Comparing MO-PPO against the hover PPO (H-PPO) algorithm reveals significant performance improvement attributable to dynamic trajectory optimization. MO-PPO effectively exploits favorable channel conditions by adapting the UAV's position, exceeding the performance of the static H-PPO approach. This underscores the critical role of trajectory optimization in maximizing the potential of ARIS-assisted CoMP-NOMA networks.

\begin{figure}[t]
    \centerline{
        \includegraphics[width=0.7\columnwidth]{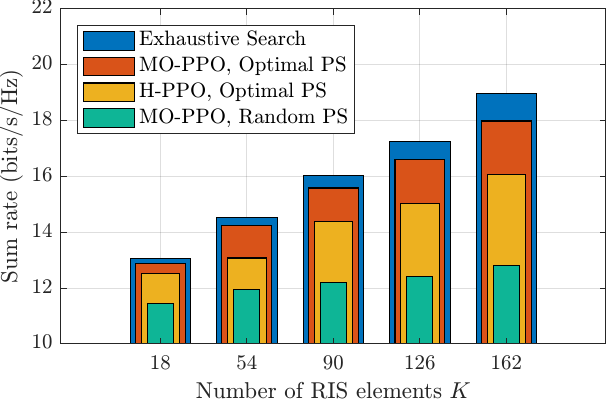}}
    \caption{Impact of the number of RIS elements on the achievable sum rate with $P_t = 10$ dBm.}
    \label{fig:bar_rate}
\end{figure}

We also analyze the impact of the number of RIS elements on the network's achievable sum rate. As shown in Fig.~\ref{fig:bar_rate}, the sum rate increases with the number of RIS elements, indicating the benefits of using larger RIS configurations for improved performance. However, a slight increase in the performance gap between the exhaustive search baseline and MO-PPO is observed as the number of RIS elements grows. This suggests the increased complexity of the action space associated with more controllable elements can subtly impact the algorithm's efficiency. Therefore, finding the optimal number of active RIS elements requires balancing performance gains with computational complexity, a challenge we aim to address in future research.

\begin{figure}[t]
    \centerline{
        \includegraphics[width=0.7\columnwidth]{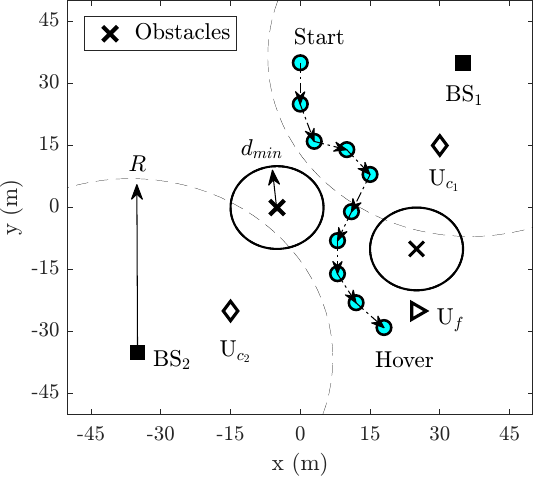}}
            \vspace{-10pt}
    \caption{Top view of the UAV trajectory obtained by the MO-PPO algorithm sampled every 25 time slots and averaged over 10 evaluation episodes}
    \label{fig:traj}
    \vspace{-10pt}
\end{figure}

Fig~\ref{fig:traj} visualizes the UAV trajectory generated by the MO-PPO algorithm. The UAV navigates around obstacles while minimizing its distance to the far user, demonstrating a careful approach to meet QoS requirements while ensuring network reliability. This behavior reflects the exploration-exploitation trade-off common in DRL algorithms, where the agent balances exploring the environment with exploiting its knowledge to maximize reward. The trajectory highlights the agent's ability to adapt to the dynamic environment and optimize network performance through effective RIS and NOMA techniques.

\section*{Acknowledgement}
The work of H. Jung was supported by IITP grant funded by the Korea government (MSIT) (No. RS-2024-00359235, Development of Ground Station Core Technology for Low Earth Orbit Cluster Satellite Communications).

\section{Conclusion}
This paper explored the potential of ARIS in enhancing CoMP-NOMA networks. We proposed a novel framework utilizing the MO-PPO algorithm to jointly optimize UAV trajectory, RIS phase shifts, and NOMA power control, aiming to maximize network sum rate while satisfying user QoS constraints. Our results demonstrated the effectiveness of MO-PPO in handling the hybrid action space and achieving near-optimal performance, showcasing the advantages of integrating ARIS, CoMP, and NOMA for future wireless networks. Although our analysis focused on a two-cell network, the proposed framework can be readily extended to accommodate a larger number of cells. Future research directions include exploring more sophisticated DRL algorithms with improved sample efficiency and convergence speed, as well as investigating the impact of imperfect CSI on the optimization process.



\bibliographystyle{ieeetr}
\bibliography{references/ref}

\end{document}